\documentclass[aps,preprint,preprintnumbers,showpacs,nofootinbib]{revtex4}
\usepackage{graphicx}

\begin{document}
\title{Neutrino mass patterns, WMAP, and Neutrinoless double beta decay}
\author{Kingman Cheung}
\email[Email:]{cheung@phys.cts.nthu.edu.tw}
\affiliation{
National Center for Theoretical Sciences, National Tsing Hua University,
Hsinchu, Taiwan, R.O.C.}
\date{\today}

\begin{abstract}
Based on the observed neutrino mass-square differences and mixings, 
we derive a few general 
patterns for the neutrino mass matrix in the flavor basis, which can correctly
describe the atmospheric and solar neutrino data.
Upon the most recent result on the absolute neutrino mass bound from WMAP,
we can rule out a large portion of the
parameter space that claimed by the positive signal in the
Heidelberg-Moscow $0\nu\beta\beta$ experiment.
In the case when one of the majorana phases is $\pi$, the effective neutrino
mass obtained is inconsistent with the  claim.
\end{abstract}
\preprint{NSC-NCTS-030301}
\maketitle

\section{Introduction}

Neutrino data collected in recent years almost established the 
hypothesis that neutrinos have masses and oscillate into each other,
especially in the recent SNO \cite{sno} and KamLAND \cite{kamland} data.
The SNO result \cite{sno} told us that the neutrino flux that 
arrived at the Earth consists of an electron neutrino component and 
a non-electron neutrino component, the sum of which matches in good
agreement with the prediction of the solar standard model without oscillation
\cite{ssm}.  The plausible explanation is that the electron neutrinos
produced at the core of the sun oscillate into other (active) neutrino flavors,
which are all detected in the SNO detector. 
On the other hand, KamLAND \cite{kamland} measured the anti-electron-neutrino
fluxes collected from a number of nuclear  plants.  With these
human-produced $\bar \nu_e$ fluxes at each plant and the initial flux known
with a high precision, the solar neutrino problem was tested for the first
time solely on Earth. 

All the combined solar neutrino data (including KamLAND data) 
now favors the oscillation of the $\nu_e$ 
into a mixture of $\nu_\mu$ and $\nu_\tau$ in the large mixing angle 
(LMA) solution \cite{solar-fits}
\begin{equation}
\label{solar}
\Delta m^2_{\rm sol} \approx 7.1 \times 10^{-5}\;\; {\rm eV}^2\,; \;\;\;
\tan^2 \theta_{\rm sol} = 0.45 \;,
\end{equation}
which gives $\sin^2 2\theta_{\rm sol} \approx 0.86$. 
The other solutions (small mixing angle, LOW, vacuum) are no longer favorable.
In addition, the sterile neutrino is also strongly disfavored.

The atmospheric neutrino data from SuperK \cite{superk} and the 
data from K2K \cite{k2k} showed a maximal mixing between the $\nu_\mu$ and
$\nu_\tau$ \cite{atm-fits} with a mass-square difference and a mixing angle:
\begin{equation}
\label{atm}
\Delta m^2_{\rm atm} \approx 2.7 \times 10^{-3}\;\; {\rm eV}^2\,, \;\;\;
\sin^2 2\theta_{\rm atm} = 1.0 \;.
\end{equation}

Yet, another piece of important information came in very recently from
WMAP \cite{wmap} 
on cosmic microwave background anisotropies.  When combined with
the data from 2dF Galaxy Redshift Survey, CBI, and ACBAR \cite{other}, WMAP is
able to pin down the amount of critical density contributed by the 
relativistic neutrinos.
Thus, it puts an upper bound on the sum of neutrino masses
\cite{weiler}
\begin{equation}
\label{71}
\sum_i m_i < 0.71 \;{\rm eV}
\end{equation}
at 95\% C.L.
Such an upper limit already contradicts the result from LSND \cite{hitoshi}.

There has been a claim of a positive signal observed in
 a neutrinoless double beta decay ($0\nu\beta\beta$) 
experiment \cite{0nbeta}.  If it is real, it will add another
constraint to the neutrino mass matrix.  The positive signal
in the $0\nu\beta\beta$ decay implies a nonzero
$m_{ee}$ entry in the neutrino mass matrix arranged in the flavor basis:
\begin{displaymath}
\left(  \begin{array}{ccc}
            m_{ee}    & m_{e\mu}  & m_{e\tau} \\
            m_{e\mu}  & m_{\mu\mu}& m_{\mu\tau}  \\
            m_{e\tau} & m_{\mu\tau} &m_{\tau\tau}  \\
       \end{array} \right ) \;.
\end{displaymath}
The 95\% allowed range of $m_{ee}$ is $0.11 - 0.56$ eV with a best value of
$0.39$ eV.
\footnote{
In Ref. \cite{0nbeta}, the allowed range of $m_{ee}$ could be widen to
$0.05-0.84$ eV if allowing a more conservative estimate on the nuclear 
matrix elements.
P. Vogel in PDG \cite{pdg} 
obtained a higher range for $m_{ee}=0.4-1.3$ eV using 
a different set of nuclear matrix elements.}
  It also implied that the electron neutrino is majorana in nature.
There are, however, arguments against this claim: see Ref. \cite{vissani,aal},
as well as counter arguments \cite{new}.

Such a large value for $m_{ee} \gg \sqrt{\Delta m^2_{\rm atm,sol}}$, if true, 
gives a nontrivial modification to the neutrino mass patterns and mixings.  
A lot of possible
mass textures that were proposed to explain the atmospheric and solar
neutrino deficits become incompatible with the new $0\nu\beta\beta$ data.
A well-known example is the Zee model \cite{zee} that can generate a bi-maximal
mixing between the $\nu_\mu$ and $\nu_\tau$ and between
the $\nu_e$ and the mixture of $\nu_\mu$ and $\nu_\tau$ \cite{jarl,paul1,otto},
however the Zee model guarantees the diagonal mass matrix elements to be zero.
Therefore, it cannot explain the nonzero $m_{ee}$ implied by the
$0\nu\beta\beta$ signal.  

More and more neutrino data are accumulated since a few years ago.  They all
contribute to the understanding of neutrino masses and mixings.  They are
very important clues to the underlying theory for neutrino masses, and perhaps
to other fermions as well.  It is therefore very important to pin down the
necessary patterns for neutrino mass matrix that can explain all the 
observations.  
The purpose of this short note is to derive the neutrino mass patterns from
the most updated solar and atmospheric neutrino fits, and the 
WMAP constraint in Eq. (\ref{71}).  Finally, we show that it is rather
marginally consistent with the claim in the $0\nu\beta\beta$ experiment.

Neutrino mass patterns under various theoretical or phenomenological 
assumptions or anatz have been considered in a lot of work \cite{xg}.

\section{Neutrino mass Patterns}

To obtain the mass pattern it is easy to work bottom-up.  We work in the basis
where the charged-lepton mass matrix is diagonal such that the mixing 
information is entirely contained in the neutrino mass matrix. From the data
we can write down specific forms of the mixing matrix $U$ and the diagonal
mass matrix $M_D$ in the mass basis, then we can obtain the
mass matrix $M_\nu$ in the flavor basis as
\begin{equation}
\label{mnu}
M_\nu = U\; M_D \;U^T \;,
\end{equation}
where the flavor basis and mass basis are related by
\begin{equation}
\left(  \begin{array}{c}
            \nu_e \\
            \nu_\mu \\
            \nu_\tau \end{array} \right ) 
  =
\left(  \begin{array}{ccc}
            U_{e1} & U_{e2} & U_{e3} \\
            U_{\mu 1} & U_{\mu 2} & U_{\mu 3} \\
            U_{\tau 1} & U_{\tau 2} & U_{\tau 3} \end{array} \right ) \;\;
\left(  \begin{array}{c}
            \nu_1 \\
            \nu_2 \\
            \nu_3 \end{array} \right ) \;.
\end{equation}
Assuming no CP violation a general form of $U$ is given by
\begin{equation}
U = U_{23} \; U_{13} \; U_{12}\;,
\end{equation}
where $U_{ij}$ is the rotation matrix about the $i$ and $j$'th mass 
eigenstates.  By considering the various data sets the general form of 
$U$ can be reduced to a simple form.
Since we know that the
atmospheric neutrino requires a maximal mixing between the $\nu_\mu$ and
$\nu_\tau$, $U_{23}$ is given by
\[
U_{23}= \left(  \begin{array}{ccc}
          1               &     0           & 0  \\
          0 &   \frac{1}{\sqrt 2} & -\frac{1}{\sqrt 2}  \\
          0 &   \frac{1}{\sqrt 2} &  \frac{1}{\sqrt 2}
            \end{array} \right ) \;.
\]
Since the angle $\theta_{13}$ is constrained by CHOOZ \cite{chooz} to be small,
we simply set $U_{13}=I$.  For the rotation between the $(1,2)$ states 
we used a generic $U_{12}$ as
\[
U_{12} = \left(  \begin{array}{ccc}
          c              &     s           & 0  \\
          -s &   c  & 0\\
          0 &    0 & 1
            \end{array} \right ) \;,
\]
where $c=\cos\theta_{12}, s=\sin\theta_{12}$. 
Therefore, $U$ is given by
\begin{equation}
\label{U}
U= \left(  \begin{array}{ccc}
          c               &     s           & 0  \\
          - \frac{s}{\sqrt 2} &   \frac{c}{\sqrt 2} & -\frac{1}{\sqrt 2}  \\
          - \frac{s}{\sqrt 2} &   \frac{c}{\sqrt 2} &  \frac{1}{\sqrt 2}
            \end{array} \right ) \;.
\end{equation}

For the mass matrix in the mass basis there are two cases: (i) normal and
(ii) inverted mass hierarchies, with a mass scale $m_0$ ($m_0$ is 
considerably larger than the atmospheric mass scale)
for the lightest one.
In the 
 normal hierarchy, the solar oscillation is between the two lighter states,
 while the atmospheric oscillation is between the heaviest and the lighter
 states.  Recall the convention that the solar oscillation is between the ``1''
 and ``2'' states, we put the diagonal mass matrix as 
\begin{equation}
 \label{md1}
 M_D^{\rm normal} = \left(  \begin{array}{ccc}
           m_0               &    0           & 0  \\
           0   &   (m_0 + \delta) \, e^{i \phi}  & 0  \\
           0   &   0             &  (m_0 + m) \, e^{i \phi'}
             \end{array} \right ) \;,
 \end{equation}
 where $\delta \simeq \Delta m^2_{\rm sol}/(2 m_0)$ ,
       $m \simeq \Delta m^2_{\rm atm}/(2 m_0)$, and we take the majorana
phases $\phi$ and $\phi'$ as 0 for the moment.  We shall later show the results
 for other values of $\phi$ and $\phi'$.
 Using Eqs. (\ref{mnu}), (\ref{U}), and (\ref{md1}), we obtain the neutrino
 mass matrix in the flavor basis:
\begin{equation}
M_\nu^{\rm normal} = 
\left(  \begin{array}{lll}
  m_0 + \delta\, s^2 & \frac{c s}{\sqrt{2}}\, \delta & 
                       \frac{c s}{\sqrt{2}}\, \delta \\
  \frac{c s}{\sqrt{2}}\, \delta   & m_0 + \frac{m}{2} + \frac{c^2}{2}\,\delta &
   -\frac{m}{2} + \frac{c^2}{2}\,\delta  \\
  \frac{c s}{\sqrt{2}}\, \delta &  -\frac{m}{2} + \frac{c^2}{2}\,\delta &
   m_0 + \frac{m}{2} + \frac{c^2}{2}\,\delta  
        \end{array}  
 \right ) \;.
\end{equation}

For the case of inverted mass hierarchy the solar neutrino oscillation is
between the two heavier states while the atmospheric neutrino oscillation
is between the lightest and the heavier states.  Recall the 
convention again that the solar oscillation is always between the ``1'' and
``2'' states, we used a $M_D^{\rm inverted}$:
\begin{equation}
 \label{md2}
 M_D^{\rm inverted} = \left(  \begin{array}{ccc}
           m_0 + m + \delta               &    0           & 0  \\
           0   &   m_0 + m  & 0  \\
           0   &   0        &  m_0 
             \end{array} \right ) \;,
 \end{equation}
in which we have chosen the phase angles to be zero.  Using Eqs. 
(\ref{mnu}), (\ref{U}), and (\ref{md2}) we obtain 
\begin{equation}
M_\nu^{\rm inverted} = 
\left(  \begin{array}{lll}
  m_0 + m+\delta\, c^2 & - \frac{c s}{\sqrt{2}}\, \delta & 
                         - \frac{c s}{\sqrt{2}}\, \delta \\
 -\frac{c s}{\sqrt{2}}\, \delta   & m_0 + \frac{m}{2} + \frac{s^2}{2}\,\delta &
   \frac{m}{2} + \frac{s^2}{2}\,\delta  \\
 -\frac{c s}{\sqrt{2}}\, \delta &  \frac{m}{2} + \frac{s^2}{2}\,\delta &
   m_0 + \frac{m}{2} + \frac{s^2}{2}\,\delta  
        \end{array}  
 \right ) \;.
\end{equation}

In general, we can use the data to fit the values of $m_0, m, \delta, 
\theta_{12}$:
\begin{eqnarray}
m_0 \approx m_{ee}\;; \quad m \simeq \Delta m^2_{\rm atm}/(2 m_0)\;; 
\nonumber \\
\delta \simeq \Delta m^2_{\rm sol}/(2 m_0)\;, \;
\theta_{12} = \theta_{\rm sol} \;.
\end{eqnarray}
We have already used a particular form of $U$ such that a maximal mixing is
always between the ``2'' and ``3'' states.

Using the WMAP constraint we obtain
\begin{eqnarray}
\sum_i m_i &\approx& 3 m_0 < 0.71 \;{\rm eV} \nonumber \\
   & \Rightarrow & m_{ee} = m_0  < 0.24 \;{\rm eV}\;.
\label{24}
\end{eqnarray}
Immediately, we can see that the $m_{ee}$ so obtained can only be in
the lower mass range of the $0\nu\beta\beta$ decay data.  In fact, the best 
fit value of 0.39 eV in the $0\nu\beta\beta$ decay is inconsistent with
the WMAP result.  

\subsection{Non-zero Majorana phases}
Next we are going to show  the results when we allow nonzero phases 
for $\phi$ and $\phi'$. For illustrations we use (ii) $\phi=\pi, \phi'=0$,
(iii) $\phi=0, \phi'=\pi$, and (iv) $\phi=\pi, \phi'=\pi$.  The case (i)
$\phi=\phi'=0$ has already been shown above.
The neutrino mass matrices in the flavor basis for normal and inverted
mass hierarchies are given, respectively, by
\widetext
\begin{eqnarray}
{\rm (ii)}\; &&\phi=\pi, \phi'=0:  \nonumber \\
&&\left(  \begin{array}{ccc}
 \cos2\theta_{12}m_0 - \delta\, s^2 & 
- \frac{\sin2\theta_{12}}{\sqrt{2}}m_0 - \frac{c s}{\sqrt{2}}\, \delta & 
- \frac{\sin2\theta_{12}}{\sqrt{2}}m_0 - \frac{c s}{\sqrt{2}}\, \delta \\
- \frac{\sin2\theta_{12}}{\sqrt{2}}m_0 - \frac{c s}{\sqrt{2}}\, \delta & 
-\frac{1}{2} \cos2\theta_{12}m_0 + \frac{m_0+m}{2} - \frac{c^2}{2}\delta & 
-\frac{1}{2} \cos2\theta_{12}m_0 - \frac{m_0+m}{2} - \frac{c^2}{2}\delta \\
- \frac{\sin2\theta_{12}}{\sqrt{2}}m_0 - \frac{c s}{\sqrt{2}}\, \delta & 
-\frac{1}{2} \cos2\theta_{12}m_0 - \frac{m_0+m}{2} - \frac{c^2}{2}\delta &
-\frac{1}{2} \cos2\theta_{12}m_0 + \frac{m_0+m}{2} - \frac{c^2}{2}\delta 
        \end{array}  
 \right )   \nonumber \\
&&
\left(  \begin{array}{ccc}
 \cos2\theta_{12}(m_0+m) +\delta\, c^2 & 
- \frac{\sin2\theta_{12}}{\sqrt{2}}(m_0+m) - \frac{c s}{\sqrt{2}}\, \delta & 
- \frac{\sin2\theta_{12}}{\sqrt{2}}(m_0+m) - \frac{c s}{\sqrt{2}}\, \delta \\
- \frac{\sin2\theta_{12}}{\sqrt{2}}(m_0+m) - \frac{c s}{\sqrt{2}}\, \delta & 
-\frac{1}{2} \cos2\theta_{12}(m_0+m) + \frac{m_0}{2} + \frac{s^2}{2}\delta & 
-\frac{1}{2} \cos2\theta_{12}(m_0+m) - \frac{m_0}{2} + \frac{s^2}{2}\delta \\
- \frac{\sin2\theta_{12}}{\sqrt{2}}(m_0+m) - \frac{c s}{\sqrt{2}}\, \delta & 
-\frac{1}{2} \cos2\theta_{12}(m_0+m) - \frac{m_0}{2} + \frac{s^2}{2}\delta &
-\frac{1}{2} \cos2\theta_{12}(m_0+m) + \frac{m_0}{2} + \frac{s^2}{2}\delta 
        \end{array}  
 \right )  \nonumber\\
{\rm (iii)}\; &&\phi=0, \phi'=\pi:  \nonumber \\
&&\left(  \begin{array}{ccc}
 m_0 + \delta\, s^2 & \frac{c s}{\sqrt{2}}\, \delta & 
                      \frac{c s}{\sqrt{2}}\, \delta \\
\frac{c s}{\sqrt{2}}\, \delta &
- \frac{m}{2} + \frac{c^2}{2}\delta & 
m_0 + \frac{m}{2} + \frac{c^2}{2}\delta \\
\frac{c s}{\sqrt{2}}\, \delta &
m_0 + \frac{m}{2} + \frac{c^2}{2}\delta &
- \frac{m}{2} + \frac{c^2}{2}\delta  \\
        \end{array}  
 \right )   \nonumber \\
&&\left(  \begin{array}{ccc}
 m_0 + m + \delta\, c^2 & -\frac{c s}{\sqrt{2}}\, \delta & 
                          -\frac{c s}{\sqrt{2}}\, \delta \\
-\frac{c s}{\sqrt{2}}\, \delta &
 \frac{m}{2} + \frac{s^2}{2}\delta & 
m_0 + \frac{m}{2} + \frac{s^2}{2}\delta \\
-\frac{c s}{\sqrt{2}}\, \delta &
m_0 + \frac{m}{2} + \frac{s^2}{2}\delta &
 \frac{m}{2} + \frac{s^2}{2}\delta  \\
        \end{array}  
 \right )   \nonumber \\
{\rm (iv)}\; &&\phi=\pi, \phi'=\pi:  \nonumber \\
&&\left(  \begin{array}{ccc}
 \cos2\theta_{12}m_0 - \delta\, s^2 & 
- \frac{\sin2\theta_{12}}{\sqrt{2}}m_0 - \frac{c s}{\sqrt{2}}\, \delta & 
- \frac{\sin2\theta_{12}}{\sqrt{2}}m_0 - \frac{c s}{\sqrt{2}}\, \delta \\
- \frac{\sin2\theta_{12}}{\sqrt{2}}m_0 - \frac{c s}{\sqrt{2}}\, \delta & 
-\frac{1}{2} \cos2\theta_{12}m_0 - \frac{m_0+m}{2} - \frac{c^2}{2}\delta & 
-\frac{1}{2} \cos2\theta_{12}m_0 + \frac{m_0+m}{2} - \frac{c^2}{2}\delta \\
- \frac{\sin2\theta_{12}}{\sqrt{2}}m_0 - \frac{c s}{\sqrt{2}}\, \delta & 
-\frac{1}{2} \cos2\theta_{12}m_0 + \frac{m_0+m}{2} - \frac{c^2}{2}\delta &
-\frac{1}{2} \cos2\theta_{12}m_0 - \frac{m_0+m}{2} - \frac{c^2}{2}\delta 
        \end{array}  
 \right )   \nonumber \\
&&
\left(  \begin{array}{ccc}
 \cos2\theta_{12}(m_0+m) +\delta\, c^2 & 
- \frac{\sin2\theta_{12}}{\sqrt{2}}(m_0+m) - \frac{c s}{\sqrt{2}}\, \delta & 
- \frac{\sin2\theta_{12}}{\sqrt{2}}(m_0+m) - \frac{c s}{\sqrt{2}}\, \delta \\
- \frac{\sin2\theta_{12}}{\sqrt{2}}(m_0+m) - \frac{c s}{\sqrt{2}}\, \delta & 
-\frac{1}{2} \cos2\theta_{12}(m_0+m) - \frac{m_0}{2} + \frac{s^2}{2}\delta & 
-\frac{1}{2} \cos2\theta_{12}(m_0+m) + \frac{m_0}{2} + \frac{s^2}{2}\delta \\
- \frac{\sin2\theta_{12}}{\sqrt{2}}(m_0+m) - \frac{c s}{\sqrt{2}}\, \delta & 
-\frac{1}{2} \cos2\theta_{12}(m_0+m) + \frac{m_0}{2} + \frac{s^2}{2}\delta &
-\frac{1}{2} \cos2\theta_{12}(m_0+m) - \frac{m_0}{2} + \frac{s^2}{2}\delta 
        \end{array}  
 \right )  \nonumber
\end{eqnarray}

\endwidetext

The present data cannot distinguish the above patterns because they all
give correct $\Delta m^2_{\rm atm}$, $\sin^2 2\theta_{\rm atm}$,
$\Delta m^2_{\rm sol}$, $\tan^2 \theta_{\rm sol}$, and $m_{ee}$ by adjusting
the parameters $m_0,m,\delta, \theta_{12}$.  The preference of one pattern 
over the others is a matter of taste.  We suggest a  fine-tuning criteria:
the stability of the maximal mixing in the atmospheric sector and 
the large mixing in the solar sector provided by the mass matrix.
The mass matrices in (i) are not stable, which is explained as follows.
Let us look at the $(2,3)$ block of the matrix, which is of the form
$\left(\begin{array}{cc}
          1 & \epsilon \\
        \epsilon & 1 \end{array} \right )$, where $\epsilon \ll 1$.  The
maximal mixing between the ``2'' and ``3'' states is very unstable.  Once
$m_{\mu\mu}$ and $m_{\tau\tau}$ differ very slightly, the maximal mixing
is destroyed.  We have verified that if $m_{\mu\mu}$ or $m_{\tau\tau}$ 
increases by 2\% the resulting $\sin^2 2\theta_{\rm atm} \sim 0.6$, already 
out of the allowed range.  
On the other hand, in case (iii) the $(2,3)$ block is of
the form
$\left(\begin{array}{cc}
           \epsilon & 1 \\
      1&  \epsilon  \end{array} \right )$.
This is a rather robust form such that even when $m_{\mu\mu}$ 
or $m_{\tau\tau}$ is changed appreciably, the maximal mixing remains.  
We have checked even if $m_{\mu\mu}$ or $m_{\tau\tau}$ increases by a factor
of two the maximal mixing solution remains for the atmospheric neutrino.
This is easy to understand.  As long as $m_{\mu\tau} \gg 
m_{\mu\mu}, m_{\tau\tau}$ we have the maximal mixing in the $(2,3)$ block.
Unfortunately, for a similar reason the matrix in case (iii) is not stable
with respect to the solar sector.  Therefore, only cases (ii) and (iv)
remain relatively 
stable with respect to both the solar and atmospheric sectors with
a small change in the matrix elements.

In cases (ii) and (iv), the effective 
mass involved in $0\nu\beta\beta$ decay is
\begin{equation}
m_{ee} \approx \; m_0 \; \cos 2 \theta_{12} \approx 0.38 m_0  < 0.09\;{\rm eV}
\end{equation}
In these favorable cases, the upper limit on
$m_{ee}$ is more stringent than that in Eq. (\ref{24}).
For this value of $m_{ee}$ it is barely inside the most conservative
95\% C.L. range of the $0\nu\beta\beta$ experiment (on the other hand,
well outside the range obtained by P. Vogel \cite{pdg}.)

For a generic majorana phase $\phi$ the effective mass $m_{ee}$ is bounded by
\begin{equation}
m_{ee} \approx m_0 \; \sqrt{ c^4 + s^4 + 2 c^2 s^2 \cos\phi} \; < \;
(0.24\;{\rm eV})\; \sqrt{ c^4 + s^4 + 2 c^2 s^2 \cos\phi} \;.
\end{equation}

\section{Conclusions}

We have derived a general neutrino mass pattern based on the most updated
fits to the atmospheric and solar neutrino data.  Using the WMAP constraint
the common mass scale $m_0$ must be less than 0.24 eV, which implies
an upper limit on the effective neutrino mass in $0\nu\beta\beta$ decay:
$m_{ee} < (0.24\;{\rm eV})
\sqrt{ \cos^4 \theta_{12} + \sin^4 \theta_{12} +2 \cos^2\theta_{12}
\sin^2 \theta_{12} \cos\phi }$, where $\phi$ is one of the majorana phases.
The largest allowed value for $m_{ee}=0.24$ eV, which is already in the
lower side of the allowed range indicated by the $0\nu\beta\beta$ 
experiment, and the best value of the experiment is already inconsistent
with the $m_{ee}$.  

In our favorable cases by considering the stability of the neutrino mass
matrix, the $m_{ee}$ is further limited to $m_{ee}<0.09$ eV, which is
just barely inside the most conservative range indicated by the experiment.
It is well below the range obtained by P. Vogel \cite{pdg} using
a different set of nuclear matrix elements. 

We conclude that the WMAP constraint has a very important impact on the
neutrino mass pattern and present a challenge to the Heidelberg-Moscow
experiment.  It also implies a sensitivity requirement on the future 
neutrinoless double beta decay experiments.

{\it Note added:} Similar conclusions on $0\nu\beta\beta$ decay have 
been obtained in Refs. \cite{weiler,hitoshi}.  Other 
work related to the WMAP result can be found in Ref. \cite{josh}.

This research was supported in part by the National Center for Theoretical
Science under a grant from the National Science Council of Taiwan R.O.C.


\end{document}